\documentstyle[12pt]{article}

\title{
On the theory of the pseudogap formation in 2D attracting 
fermion systems } 
\author{{\sl V.M.~Loktev}\\ 
{\sl  Bogolyubov Institute for Theoretical Physics,}\\ 
{\sl Metrologichna str. 14-b, Kyiv, 252143 Ukraine}\\ 
{\sl and V.M.~Turkowski}\\
{\sl Shevchenko Kyiv University}\\ 
{\sl Acad. Glushkova prosp. 6, Kyiv, 252127 Ukraine}} 
\date{}
\begin{document} \maketitle


\begin{abstract}
Two-dimensional system of the fermions 
with the indirect Einstein 
phonon-exchange attraction and added local four-fermion interaction 
is considered. It is shown that in such a system at resulting
attraction between particles a new nonsuperconducting phase arises 
along with the normal and superconducting phases. In this, called 
"abnormal normal", or pseudogap, phase the absolute value  
of the order parameter is finite but its phase is a random quantity.  
It is important that the new phase really exists at low carrier 
density only, i.e. it shrinks with doping increasing in the case of 
phonon attraction. The relevance of the 
results for high-temperature superconductors is speculated.  
\end{abstract}

{\em PACS:} 67.20.+k, 74.20.-z, 74.25.-q, 74.72.-h

\eject

\section{Introduction}

The problem of an adequate description of the physical properties of 
high-temperature superconductors (HTSCs) still remains one of the 
actual problems of the modern solid state physics. It is
connected with some peculiar properties of HTSCs. Among them there 
are such as quasi-2D character of electronic (and magnetic) 
properties, a relatively low and changeable carrier density $n_f$
and its influence on properties of HTSCs (see, 
for example, review \cite{Lok1}).  

Nowadays, one of the widely discussed 
questions on HTSCs is the problem of the so called 
pseudogap (or spin gap if magnetic subsystem of HTSCs is taken into
account) \cite{Mar1,Din1,Loes1}, which is usually experimentally 
observed as a loss in the spectral weight of quasiparticle (or spin) 
excitations in normal state samples with lowered carrier density 
$n_f$ \cite{Lev1,Bat1,Ong1}. Corresponding underdoped samples reveal 
some specific spectral, magnetic and thermodynamic pecularities 
which still continue to be not sufficiantly understood now. 
Moreover, the striking difference between the low (underdoped) and 
high (overdoped) density regions in HTSCs is increasingly debated and 
is considered as one of the very central and key questions in physics 
of the cuprates \cite{Ber1, Pin1}.

The possibility of experimental changing of $n_f$ value in HTSCs puts
a rather general theoretical problem of the description of the 
crossover from composite boson superfluidity (low $n_f$) to Cooper 
pairing (large $n_f$) when $n_f$ increases (in other words, a 
description of the transition from the so called underdoped regime 
to the overdoped one).  Such a crossover was already studied for  
3D and quasi-2D systems (see reviews 
\cite{Ran1,Gor1}).  2D case has been considered for the present at 
temperature $T=0$ only \cite{Ran1,Gor2} what is connected with the 
Hohenberg-Mermin-Wagner theorem which forbids any 
homogeneous (long-range) order in pure 2D systems at $T\not= 0$ due 
to the long-wave fluctuations of the charged order parameter (OP).

The problem of the inhomogeneous condensate 
(Berezinskii-Kosterlitz-Thouless, or BKT, phase) formation was
also considered despite of some 
difficulties in 2+1 relativistic field models \cite{Mac1} where 
the fermion concentration effects are irrelevant. At the same time 
these effects were studied in nonrelativistic model in  \cite{Drec1}, 
for example, without taking into account the existence of the neutral 
OP $\rho$. Its consideration proves to be very 
important (see \cite{Gus1}) and results in the formation of a 
separate equilibrium phase with $\rho\not= 0$ which is located 
on the phase diagram of a system between 
normal and superconducting (here - BKT) ones. Due to fluctuations of 
the OP phase this new state of a system is of course also 
non-superconducting.

In this paper an attempt is made to study the crossover as well as 
the above mentioned new phase formation possibility in 2D fermion 
system with both a more realistic indirect (phonon) and also a direct 
(local) four-fermion (4F-) interactions. Thus, the work is to a 
certain extent a specific and non-trivial generalization of 
the preliminary short communication \cite{Gus1} where this 
non-superconducting phase appearance was studied for 4F-case only 
and of the paper \cite{Lok2} where Fr\"ohlich model was used for the 
investigation of the crossover at $T=0$. As it will be seen in the 
boson exchange model (in contrast to the pure 4F-case), the new phase 
really exists when $n_f$ is rather small what allows to relate this 
result to underdoped HTSC compounds. But actually it is 
interesting to take into account a more real situation with 
an indirect attraction and some kind of local repulsion which may 
correspond to the short-range (screened) Coulomb interaction 
between carriers. In general case we, however, suppose that 
4F-interaction can be repulsive and attractive as well. Besides, the 
case of total repulsion allows to explore the 
fermion-antifermion (electron-hole) pairing channel which in spite
physical difference can be formally described by the same manner. 

\section{Model and main equations}

Let us choose the simplest Hamiltonian density in the form:  
\begin{eqnarray} 
H(x) =-\psi_{\sigma}^{\dagger}(x) 
      \left(\frac{\nabla^2}{2m}-\mu\right)\psi_{\sigma}(x) + 
      H_{ph}(\varphi (x))+\nonumber \\ 
      g_{ph}\psi_{\sigma}^{\dagger}(x)\psi_{\sigma}(x)\varphi (x)- 
      g_{4F}\psi_{\uparrow}^{\dagger}(x)\psi_{\downarrow}^{\dagger}(x) 
      \psi_{\downarrow}(x)\psi_{\uparrow}(x),\ \  (x=\mbox{\bf 
      r},t), \label{1} 
 \end{eqnarray} 
where $\psi_{\sigma}(x)$ is a 
fermionic field with an effective mass $m$ and spin $\sigma =\uparrow 
,\downarrow $; $\mu$ is the chemical potential of the fermions which 
fixes $n_{f}$; $\varphi (x)$ is a phonon field operator, $g_{ph}$ and 
$g_{4F}$ are the electron-phonon and the 4F-interaction coupling 
constants, respectively.  As we said above, $g_{4F}$ can be 
positive (fermion-fermion attraction) or negative 
(fermion-antifermion attraction); in (\ref{1}) we set
$\hbar=k_B=1$.

In (\ref{1}) $H_{ph}$ is the Hamiltonian of free phonons which 
can be described by the propagator  
\begin{equation} 
D(i\Omega_n ) 
=-\frac{\omega_0^2}{\Omega_n^2+\omega_0^2} , \label{2}
\end{equation}
where $\Omega_n=2n\pi T$ ($n$ is an integer) is the Matsubara frequency
\cite{Abr1}. As it follows from (\ref{2}), the propagator 
$D(i\Omega_n)$ has been chosen in the simplest form with 
$\omega_0$ being the Einstein (dispersionless) phonon frequency. It 
was done because of several reasons: first, this propagator 
gives a possibility to integrate the equations obtained; 
second, it is precisely the optic phonon and quadrupolar exciton modes 
with their relatively weak dispersion are widely considered as 
exchange bosons that can contribute into the hole-hole attraction in 
HTSCs \cite{Lok1,Pash,Ginz}, and third, the qualitative results 
concerning retardation effects do not strongly depend on the model 
studied. But on the other hand, the propagator (\ref{2}) for the 
model under consideration can hardly be used for quantitative 
description of the cuprates and also their spin-wave branches which, 
as it is well-known, have linear dispersion law.

It is important that the Hamiltonian (\ref{1}) is 
invariant under symmetry transformations of two types \cite{Nambu},
namely: 
\begin{equation} 
\psi_{\sigma}(x)\rightarrow\psi_{\sigma}(x)e^{i\alpha (x)},\ \ 
   \psi_{\sigma}^{\dagger}(x)\rightarrow\psi_{\sigma}^{\dagger}(x)
   e^{-i\alpha (x)}\label{s1}
\end{equation}
and 
$$
\psi_{\uparrow}(x)\rightarrow\psi_{\uparrow}(x)e^{i\alpha (x)},\ \
   \psi_{\downarrow}(x)\rightarrow\psi_{\downarrow}(x)e^{-i\alpha (x)}, 
$$ 
\begin{equation} 
   \psi_{\uparrow}^{\dagger}(x)\rightarrow\psi_{\uparrow}^{\dagger}(x)
    e^{-i\alpha (x)},
    \psi_{\downarrow}^{\dagger}(x)\rightarrow\psi_{\downarrow}^{\dagger}(x)
    e^{i\alpha (x)}\label{s2}
\end{equation}
which must be taken into account. The phase $\alpha (x)$ in (\ref{s1})
and (\ref{s2}) is real.

With the purpose to calculate the phase diagram of the system it is 
necessary to find its thermodynamic potential. It can be calculated
by making use of the auxiliary bilocal field method (see, for example,
\cite{Kug1}), which is a generalization of the standard
Hubbard-Stratonovich one for the boson-exchange case. Then the grand
partition function $Z$ can be expressed through a path integral 
over the fermionic $\psi_{\sigma}(x)$ and the complex auxiliary 
fields (for example, $\phi (x,x')\sim  \    \ 
<\psi_{\uparrow}^{\dagger}(x)\psi_{\downarrow}^{\dagger}(x')>$). 

In the case of the model (\ref{1}) it is convenient following
the Ref.\cite{Von1} to introduce the bispinor 
\begin{equation} 
\Psi^{\dagger}(x)=(\psi_{\uparrow}^{\dagger}(x),\psi_{\downarrow}
^{\dagger}(x),\psi_{\uparrow}(x),\psi_{\downarrow}(x))\label{3}
\end{equation}
and its hermitian conjugate one which here are the analogous of the
Nambu spinors \cite{Sch1}. After the substitution of (\ref{3}) in 
(\ref{1}) the Hamiltonian takes the form:  
\begin{eqnarray} 
H(x)=-\Psi^{\dagger}(x)(\frac{\Delta^2}{2m}+\mu)I\otimes\tau_{z}\Psi 
(x)- g_{ph}\Psi^{\dagger}(x)I\otimes\tau_{z}\Psi (x)\varphi 
(x)-\nonumber \\ g_{4F}\Psi^{\dagger}(x)I\otimes\tau_{z}\Psi 
(x)\Psi^{\dagger}(x)I\otimes\tau_{z} \Psi (x)+\varphi 
(x)D^{-1}(x)\varphi (x),\label{4} 
\end{eqnarray} 
where $I\otimes\tau_{z}$ is 
the direct product of the unit $I$ and Pauli $\tau_{z}$ $2\otimes 2$ 
matrices; $D(x)$ is defined by (\ref{2}). In such a representation of 
the Hamiltonian (\ref{4}) and the field variables (\ref{3}) the 
Feinman diagram technique becomes applicable in its 
ordinary form \cite{Von1}.  Thus, after standard excluding of the 
boson field $\varphi (x)$, the Lagrangian of the system can be 
expressed by the formula:  
\begin{eqnarray} 
L(x_1,y_1,x_2,y_2)=\Psi^{\dagger}(x)[-\partial_{\tau}+
(\frac{\Delta^{2}}{2m}+\mu ) I\otimes\tau_{z}]\Psi (x)-\nonumber \\ 
\frac{1}{2}\Psi (x_1)\Psi^{\dagger}(y_1)I\otimes\tau_{z} 
K(x_1,y_1;x_2,y_2)\Psi 
(x_2)\Psi^{\dagger}(y_2)I\otimes\tau_{z}.\label{5} 
\end{eqnarray} 
The kernel $K$ is the effective non-local inter-particle interaction 
function and will be explicitly defined in the momentum 
space below. 

In order to explore the pairing possibility in the 
system let us introduce the bilocal auxiliary field, 
or OP, 
\begin{equation} 
\phi (x_1,y_1)=\tau_{z}K(x_1,y_1;x_2,y_2)\Psi 
(x_2)\Psi^{\dagger}(y_2)\equiv 
iI\otimes\tau_y\phi_{ch}(x_1,y_1)+\tau_{x}\otimes\tau_{x}
\phi_{ins}(x_1,y_1) 
\label{6} 
\end{equation} 
(the integration over $x_2$ and $y_2$ is assumed). Here 
$\phi_{ch}\sim 
<\psi_{\uparrow}^{\dagger}\psi_{\downarrow}^{\dagger}>$ and 
$\phi_{ins}\sim 
<\psi_{\downarrow}^{\dagger}\psi_{\uparrow}>$ are 
electron-electron (charged) and electron-hole (insulating) 
OP, respectively (we neglect non-zero spin pairing).  The auxiliary 
fields $\phi_{ch}$ and $\phi_{ins}$ are responsible for the dynamical 
symmetry breaking (in according with (\ref{s1}) and (\ref{s2}), 
correspondingly).

Adding to (\ref{5}) a zero term  
$$
\frac{1}{2}[\phi (x_1,y_1)-K(x_1,y_1;x'_1,y'_1)\Psi 
(x'_1)\Psi^{\dagger}(y'_1)]K^{-1}(x_1,y_1;x_2,y_2)
[\phi (x_2,y_2)- 
$$ 
$$ 
K(x_2,y_2;x'_2,y'_2)\Psi (x'_2)\Psi^{\dagger} (y'_2)] 
$$ 
to cancel the 4F-interaction, one could obtain the Lagrangian in the 
form:  
\begin{eqnarray} 
L(x_1,y_1;x_2,y_2)=\Psi^{\dagger}(x_1)
[-\partial_{\tau}+(\frac{\Delta^2}{2m}+
\mu )I\otimes\tau_{z}- \phi (x_1,y_1)]\Psi (y_1)+\nonumber\\
\frac{1}{2}\phi 
(x_1,y_1)K^{-1}(x_1,y_1;x_2,y_2)\phi (x_2,y_2),\label{7} 
\end{eqnarray}

Let us transform the expression for the kernel $K$; then in the 
momentum space it is
$$ 
K(x_1,y_1;x_2,y_2)= 
$$ 
$$ 
\int\frac{d^3Pd^3p_1d^3p_2}{(2\pi)^9}K_P(q_1;q_2)\\
exp\left[ -iP(\frac{x_1+y_1}{2}-\frac{x_2+y_2}{2})-
ip_1(x_1-y_1)-ip_2(x_2-y_2)\right]
$$
where $p_i=(\vec{p_i},\omega_i) (i=1,2)$ and 
$P=(\vec{P},\omega )$ designate the relative and the centre of mass 
momenta, respectively. By the definition the kernel 
$K_P(p_1;p_2)$ is in fact independent of $P$ (so we omit index $P$ 
henceforth) and aquires the simple form
\begin{equation} 
K(p_1;p_2)=g_{ph}^2D(p_1-p_2)-g_{4F}\label{71}  
\end{equation} 
which will be used in (\ref{7}). The last expression 
evidently demonstrates that the total character of the effective 
inter-particle interaction as it always takes place in such a 
situation \cite{Geil,Sch1} defined by the possible competition between 
the first (retarded) and second (retardless) terms in (\ref{71}),
or their common action.
 
The partition function can be written as:  
$$ 
Z=\int {\cal D}\Psi^{\dagger} {\cal D}\Psi {\cal D}\phi {\cal D}\phi 
^{*} \exp\left[-\beta\int L(\Psi^{\dagger},\Psi ,\phi^{*},\phi 
)dxdy\right] 
$$
$$
\equiv \int{\cal D}\phi {\cal 
D}\phi^{*}\exp(-\beta\Omega [{\cal G}]), \ \ (\beta =1/T), 
$$ 
where $\Omega [{\cal G}]$ is the thermodynamic potential which in the 
"leading order" is 
\begin{equation} 
\beta\Omega [{\cal G}]=- \mbox{Tr} \left[Ln {\cal 
G}^{-1}+\frac{1}{2}\mbox{Tr}(\phi K^{-1}\phi )\right], \label{9} 
\end{equation} 
where Tr includes 2D spatial $\mbox{\bf r}$ and 
"time" $0\leq\tau\leq\beta$ integrations as well as the 
standard trace operation. The full Green function of a system is 
\begin{equation} 
{\cal G}^{-1}  = 
-\partial_{\tau}+\tau_z\left(\frac{\nabla^2}{2m}+\mu\right)I\otimes\tau_{z}+
\phi .  
\label{10} 
\end{equation} 
From (\ref{9}) and (\ref{10}) we arrive 
to the $\phi$-equation (the Schwinger-Dyson one):  
\begin{equation} 
\delta\Omega /\delta\phi 
=\phi-\int\frac{d^2\vec{k}d\omega}{(2\pi )^3}K(p;\vec{k},\omega ){\cal 
G}(\vec{k},\omega )=0.\label{11} 
\end{equation} 
Substituting 
(\ref{11}) into (\ref{9}) one can obtain the expression for $\Omega 
({\cal G})$:  
$$ 
\beta\Omega ({\cal G}) =-\mbox{Tr}\mbox{Ln}{\cal 
G}^{-1} +\frac{1}{2}\mbox{Tr}{\cal G}K{\cal G}, 
$$ 
The last is the well-known Cornwell-Jackiw-Tomboulis formula for the
effective action in the one-loop approximation \cite{Cor1}. Using
(\ref{11}) we can rewrite this expression in the form  
\begin{equation} 
\beta\Omega ({\cal G 
})=-\mbox{Tr}[\mbox{Ln}{\cal G}+ \frac{1}{2}[{\cal G}{\cal 
G}_0^{-1}-1]].\label{12} 
\end{equation} 

As it was shown by Thouless et al.\cite{Tau1} (see also \cite{Gus1})
in 2D case it is natural to pass to a new parametrization of the OP 
(\ref{6}) - its absolute value and the phase, namely:  
$$ 
\phi_{ch}(x,y)=\rho_{ch}(x,y)exp[-i(\theta (x)+\theta 
(y))],
$$ 
\begin{equation} 
\phi_{ins}(x,y)=\rho_{ins}(x,y)exp[-i(\theta (x)+\theta 
(y))],\label{13} 
\end{equation}
where $\rho_{ch}$ and $\rho_{ins}$ are real.

As it will be shown below, with the given kernel (\ref{71})
there can arise only one ($\phi_{ch}$ or $\phi_{ins}$) OP. Therefore, 
it is necessary to make, simultaneously with (\ref{13}), the 
spinor transformation (in according with (\ref{s1}) and (\ref{s2})) 
\begin{equation} 
\Psi^{\dagger}(x)=\chi^{\dagger}(x)exp(i\theta 
(x)I\otimes\tau_{z}),\label{141} 
\end{equation} 
\begin{equation}
\Psi^{\dagger}(x)=\chi^{\dagger}(x)exp(i\theta (x)I\otimes\tau_{z}),
\label{142}
\end{equation}
(the spinor $\chi (x)$ is real and formally corresponds to 
chargeless fermions).  It is easy to see from (\ref{13}), 
(\ref{141}) and (\ref{142}) that the phase dependences of the charged 
and insulating OPs are similar.  Below we shall obtain 
$\theta$-corrections for the $\phi_{ch}$ case, because the final 
equations for $\phi_{ins}$ will be the same up to substitution 
$\rho_{ch}\rightarrow\rho_{ins}$. The reason is that when $K(p_1,p_2)$
describes the attraction (charge pairing channel) the symmetry of the 
Lagrangian under operations (\ref{s1}) proves to be crucial for the 
representation (\ref{141}); while when $K(p_1,p_2)$ corresponds to 
the repulsion (chargeless, or insulating, pairing channel) the 
symmetry (\ref{s2}) becomes already important and the representation 
(\ref{142}) must be used as a "working" one. With this difference the 
rest of the calculations are almost identical and so we 
shall consider in detail the charge channel which is most 
interesting for metallic (superconducting) systems.
 
In variables (\ref{141}) the Green function (\ref{10}) 
transformes to  
\begin{eqnarray} {\cal G}^{-1} & = & 
-\partial_{\tau}+I\otimes\tau_z\left(\frac{\nabla^2}{2m}+\mu\right)+ 
iI\tau_y\rho_{ch}+  \nonumber \\ && 
I\otimes \tau_z\left(\partial_{\tau}\theta+ 
\frac{\nabla\theta^2}{2m}\right)+iI\otimes 
I\left(\frac{\nabla^2\theta}{2m}+ 
\frac{\nabla\theta\nabla}{m}\right)\equiv G^{-1}(\rho_{ch}) -\Sigma 
(\partial\theta ).  \label{15} \end{eqnarray} 

Then using (\ref{15}) supposing $\theta$ gradients are small (the 
hydrodynamic approximation) and taking them into account up to the 
second order the effective potential (\ref{12}) can be naturally 
divided it two parts:  $\Omega =\Omega_{kin}(\rho_{ch},\nabla\theta ) 
+ \Omega_{pot}(\rho_{ch})$ where in $(\nabla \theta)^{2}$ 
approximation 
\begin{eqnarray} 
\beta\Omega_{kin}(\rho_{ch},\nabla\theta ) 
=\mbox{Tr}\left[G\Sigma -G_0\Sigma+\frac{1}{2}G\Sigma 
G\Sigma - \right. \nonumber \\
\left. \frac{1}{2}G_0\Sigma G_0\Sigma + 
\tau_{x}\otimes I\frac{1}{2}i\rho_{ch} G(G \Sigma + G \Sigma 
G \Sigma) \right].\label{16} 
\end{eqnarray} 
Assuming now that 
$\rho_{ch}(x,y)$ is homogeneous \footnote{Equations for $\rho_{ch}$ 
and $\rho_{ins}$ will be obtained below and, as was shown in 
\cite{Lok2}, it is an admissible approximation to put in them 
the value $\rho_{ch}$ (and $\rho_{ins}$) independent of spatial and 
time variables.} after somewhat tedious but otherwise straightforward 
calculation one can obtain from (\ref{16}):
\begin{equation} 
\Omega_{kin}(\rho_{ch},\Delta\theta )=
\frac{T}{2}\int_{0}^{\beta}d\tau\int d^2\mbox{\bf r} J(\mu, T, 
\rho_{ch}(\mu ,T))(\nabla\theta)^2, \label{17} 
\end{equation} 
where 
$$ J(\mu ,T,\rho_{ch}(\mu ,T))= 
$$ 
$$ 
\frac{1}{2\pi}(\sqrt{\mu^2+\rho_{ch}^2}+\mu + 
2T\ln\left[1+\exp\left(-\frac{\sqrt{\mu^2+\rho_{ch}^2
}}{T}\right)\right]-
$$
\begin{equation}                                               
\frac{T}{\pi}\left[1- 
\frac{\rho_{ch}^2}{4T^2} \frac{\partial}{\partial 
(\rho_{ch}^2/4T^2)}\right]\int_{-\mu 
/2T}^{\infty}dx\frac{x+\mu /2T} 
{\cosh^2\sqrt{x^2+\rho_{ch}^2/4T^2}} \label{18} 
\end{equation}
plays the role of neutral OP stiffness.
Note that in comparison with the retardation free 4F-model \cite{Gus1}
the last expression contains one more term, namely: the term with 
the derivative.

The equation for the temperature $T_{BKT}$ of the BKT transition can 
be written down after direct comparison of the kinetic term  
(\ref{17}) in the effective action with the Hamiltonian of the 2D
XY-model which has the formally identical form \cite{Izum1}. Hence it 
easy to conclude that 
\begin{equation} 
\frac{\pi}{2}J(\mu ,T_{BKT},\rho_{ch} (\mu ,T_{BKT})) = T_{BKT}.  \label{19} 
\end{equation} 
The essential difference of this equation obtained from that for the
XY-model is its spontaneous dependence on $\mu$ (or $n_f$) and 
$\rho_{ch}$.

To complete the set of self-consitent equations which 
allow to trace an explicit dependence of $T_{BKT}$ on $n_{f}$, the 
equations for $\rho_{ch}$ and $\mu$ have also to be 
given. In particular, the equation for $\rho_{ch}(i \omega_{n})$  
is nothing else but $(\ref{11})$ with 
$\nabla \theta = 0$, i.e. the Green function $G$ of the neutral 
fermions substitutes $\cal G$, so that $(\ref{11})$ in 
frequency-momentum represantation takes the form 
\begin{equation} 
\left(
\begin{array}{l}
\rho_{ch}(i \omega_{n}) \\
\rho_{ins}(i \omega_{n})
\end{array}\right)
 = 
T\sum_{m=-\infty}^{\infty}\int\frac{d^2\mbox{\bf 
k}}{(2\pi)^2}
\left(
\begin{array}{l}
-\rho_{ch}(i\omega_m)\\
+\rho_{ins}(i\omega_{m})
\end{array}\right)
\frac{K(\omega_n,\omega_m)}
{\omega_m^2+\varepsilon^2 
(\mbox{\bf k})+\rho_{ch}^2 
(i\omega_m)+\rho_{ins}^2(i\omega_m)},
\label{20} 
\end{equation} 
where $\omega_n=(2n+1)\pi T$ is the Matsubara fermionic 
frequency \cite{Sch1} and kernel $K(\omega_m,\omega_n)$ is defined 
above. We cited the final equations for both OPs, $\rho_{ch}$ and
$\rho_{ins}$, in order to show only that they indeed are the same 
but alternative if the kernel $K$ changes the sign.

An analitical solution of these equations, as well as obtaining both
the equation (\ref{19}) and the number equation needed is only 
possible if one supposes that $\rho_{ch}(i\omega_n)$ does not depend 
on the Matsubara frequencies (see footnote on p.8.).

Making use of this approximation the number equation 
which follows from the ordinary condition $V^{-1}\partial\Omega[{\cal 
G}] /\partial\mu =-n_{f}$ ($V$ is a volume of a system) and 
is crucial for crossover description has to be added to (\ref{19}) 
 and (\ref{20}) for self-consistency; so one comes to 
\begin{equation} 
\sqrt{\mu^2+\rho_{ch}^2}+\mu 
+2T\ln\left[1+\exp 
\left(-\frac{\sqrt{\mu^2+\rho_{ch}^2}}{T}\right)\right]= 
2\epsilon_F, \label{21} 
\end{equation} 
where $\epsilon_F=\pi n_f/m$ 
is the Fermi energy of free 2D fermions with the simplest quadratic 
dispersion law.  Thus, in the case under consideration all unknown 
quantities $\rho_{ch}, \mu$ and $T_{BKT}$ are the explicit
functions of $n_f$. 

\section{Analysis of the solutions}

Unlike the usual (with $T$-independed unit vector) 
XY-model, in the superconducting one
there exist two critical temperatures:
$T_{\rho}$ where formally the complete OP given by (\ref{6})
arises but its phase is a random quantity, i.e.  $<\phi (x,y)>=0$
\footnote{Because of $\rho_{ch}$ and $\rho_{ins}$ can not exist 
simultaneously (see (\ref{20})) the index $\rho$ means the only OP, 
which appears at finite sign of total interaction.}, and another one, 
$T_{BKT}<T_{\rho}$, where the phase of the OP becomes ordered, so 
that $<\phi (x,y)>\not=0$.  In other words, the temperature 
$T_{\rho}$ is in fact the temperature of appearance of neutral OP 
only which has discrete symmetry and thus is not at variance with 
general theorems.  Recall, that according to the equations obtained 
above both these temperatures, what is important, directly depend on 
the carrier density in the system.  
  
Critical temperature $T_{\rho}$ can be found from 
(\ref{18})-(\ref{21}) by putting $\rho_{ch}=0$ (what in 
accordance with derivation of these 
equations corresponds, in fact, to the mean-field approximation). As 
a result, a 2D metal with temperature decreasing passes from normal 
phase ($T>T_{\rho}$) to another one where averaged homogeneous 
(charged) OP $<\phi (x,y)>=0$, or, what is the same, 
superconductivity is absent, but chargeless OP $\rho_{ch}\not= 0$.  
It is evident that the pseudogap is formed just in the temperature 
region $T_{BKT}<T<T_{\rho}$, because, as follows from the above 
formulas (see, for instance, (\ref{18})-(\ref{21})), 
$\rho_{ch}=\rho_{ch}(T)$ enters all spectral 
characteristics of 2D metal in the same way as the superconducting 
gap $\Delta (T)$ enters into corresponding expressions for ordinary 
superconductors.  It justifies why this new phase can be called the 
"abnormal normal" phase or better pseudogap one.  The density of 
states near $\epsilon_F$ in the pseudogap is definitely less than in 
the normal phase, but is not to be equal zero as in superconducting 
one. The latter has to be checked by direct calculation of the 
one-particle fermion Green function what is most likely a separate
problem which we do not touched upon here.

The phase diagram of the 
system can be found from the equations (\ref{18})-(\ref{21}). 
There are different behaviours of $T_{\rho }(n_f)$ and $T_{BKT}(n_f)$ 
for various correlation between interaction constants. 

1) $g_{4F}>0, g_{ph}=0$ (retardless interaction).

This case has been partly analyzed in Ref.\cite{Gus1}. It 
corresponds to fermion-fermion pairing due to the local interaction. 
Note (see (\ref{20})) that in the case of resulting attraction 
between fermions  
fermion-antifermion (or electron-hole) insulating pairing channel is 
absent, i.e. $\rho_{ins}=0$. The phase diagram for this case is 
presented in fig.1.  It shows that the abnormal normal phase 
exists at any concentration value $n_f$ and the temperature width of 
this phase region weakly increases with $n_f$ increasing, and BKT 
phase always begin to form when $\rho_{ch}(T_{BKT})$ is finite.

At $\epsilon_{F}\rightarrow 0$ the temperature of BKT phase formation 
is defined by equality $T_{BKT}=\epsilon_{f}/2$, and $T_{\rho}$ as 
function of $n_{f}$
can be found from the equation $T_{\rho}\ln 
(T_{\rho}/\epsilon_{F})=W\exp (-2/g_{4F}m)$, which follows from 
(\ref{20}) ($W$ is the conduction band width). 

2) $g_{4F}>0, g_{ph}\not= 0$.

The situation here is almost the same as previous one. The 
presence of the indirect interaction leads 
to the effective growth of the effective 4F-interaction constant 
$g_{4F}^{*}>g_{4F}$ (see (\ref{20})).  The latter in one's 
turn simply results in increasing of the region between 
$T_{\rho}$ and $T_{BKT}$ (fig.2) and keeps the form shown on fig.1. 

3) $g_{4F}=0, g_{ph}\not= 0$ (a pure indirect interaction).

This is one of the most interesting cases because it corresponds
to the widely accepted electron-phonon (or BCS-Bogolyubov-Eliashberg) 
model of superconductivity. The numerical calculations of the phase 
diagram is presented in fig.2. This diagram shows 
that comparatively large region with the abnormal normal 
(pseudogap) phase exists at rather low carrier concentrations only 
and its temperature area shrinks at $n_f\rightarrow\infty$. Such a 
behaviour qualitatively agrees with that one
which takes place in real HTSCs samples 
\cite{Lev1,Bat1,Ong1,Ber1} demonstrating that pseudogap (and spin gap 
also) region is observed in underdoped samples.

Indeed, it is not difficult to make certain that the 
assymptotics for $T_{\rho }(n_f)$ and $T_{BKT}(n_f)$ have the 
following forms:

i) when ratio $\epsilon_F/\omega_0\ll 1$ (very low fermion
density, or local pair case) the first of them satisfies the 
equation $T_{\rho}\ln (T_{\rho}/\epsilon_F)= \omega_0\exp 
(-4\pi/g_{ph}^{2}m)$ which immediatedly results in $\partial 
T_{\rho}(n_f)/ \partial n_f |_{n_f\rightarrow 0}\rightarrow\infty$.  
At the same time the temperature $T_{BKT}$ at $n_{f}\rightarrow 0$ 
has another carrier density dependence and as above 
$T_{BKT}=\epsilon_F/2$ what simply means that here it again is 
equal to the number of composite bosons; in this density region 
$T_{\rho}/T_{BKT}\gg 1$ (such an inequality is also correct 
for the pure 4F-interaction).

ii) in the case $\epsilon_F/\omega_0\gg 1$ (very 
large fermion density, or Cooper pair case) one easily arrives 
to the standard BCS value:  $T_{\rho}=(2\gamma\omega_0/\pi) \exp 
(-2\pi/g_{ph}^2m)\equiv T_{BCS}^{MF}=(2\gamma 
/\pi)\Delta_{BCS}$ ($\Delta_{BCS}$ is the usual one-particle 
BCS gap at $T=0$). In other words, the temperature 
$T_{\rho}$ in this limit becomes equal to its BCS value 
\footnote{Being equal (in 
mean field approximation only) these temperatures ($T_{\rho}$ and 
$T_{BCS}^{MF}$) are in fact different: if $T_{BCS}^{MF}$ immediately 
falls down to zero as fluctuations are taken into account, $T_{\rho}$ 
does not and is renormalized only.}.  
The $T_{BKT}$ asymptotics for this 
case is not so evident and requires more detailed consideration. 

First of all, it is naturally to 
suppose that for large $n_f$ value $T_{BKT}\rightarrow 
T_{\rho}$.  Then it is necessary to check the dependence of 
$\rho$ on $T$ as $T\rightarrow T_{\rho}$.  For that the 
equation (\ref{20}) can be transformed to:  
$$ 
\frac{2\pi}{g_{ph}^2m} 
=\int_{0}^{\infty}dx\left(\frac{\tanh\sqrt{x^2+\rho_{ch}^2/4T^2}} 
{\sqrt{x^2+\rho_{ch}^2/4T^2}}-\frac{\tanh\sqrt{x^2+\rho_{ch}^2/4T^2}-
\tanh 
(\omega_0/2T)}{2(\sqrt{x^2+\rho_{ch}^2/4T^2}-\omega_0/2T)}-\right.  
$$ \begin{equation} 
\left.\frac{\tanh\sqrt{x^2+\rho_{ch}^2/4T^2}+\tanh 
(\omega_0/2T)} {2(\sqrt{x^2+\rho_{ch}^2/4T^2}+\omega_0/2T)}\right) 
\label{22} 
\end{equation} 
(where it was used that in this concentration region, 
the ratio $\mu /2T_{\rho}\simeq\epsilon_F/2T_{\rho} \gg 1$ because 
of $\mu\simeq\epsilon_{F}$\cite{Ran1,Gor1,Gor2,Lok2}).

On account of usually $\omega_0/2T_{\rho}\gg 1$, only very 
small $x$ give the main contribution to the integral (\ref{22}) (it 
is seen from the limit $\rho /2T_{\rho} \rightarrow 0$ when 
 $\epsilon_F/\omega_0\rightarrow\infty$). Therefore the latter 
 expression takes the approximate form:  
\begin{equation} 
\frac{2\pi}{g_{ph}^2m}=\int_{0}^{\infty}dx\left(\frac{\tanh\sqrt{x^2+
\rho_{ch}^2/4T^2}}
{\sqrt{x^2+\rho_{ch}^2/4T^2}}-\frac{1}{x+\omega_0/2T}\right).\label{23} 
\end{equation}
 On the other hand, the condition $\rho_{ch}=0$ in (\ref{23}) leads to
the equation
\begin{equation} 
\frac{2\pi}{g_{ph}^2m}=\int_{0}^{\infty}dx\left(\frac{\tanh 
x}{x}- \frac{1}{x+\omega_0/2T_{\rho_{ch}}}\right).\label{24} 
\end{equation}
for $T_{\rho}$.
From (\ref{23}) and (\ref{24}) it directly follows that 
$$
\int_{0}^{\infty}dx\left(\frac{\tanh 
x}{x}-\frac{\tanh\sqrt{x^2+\rho_{ch}^2/4T^2}} 
{\sqrt{x^2+\rho_{ch}^2/4T^2}}\right)=\ln\frac{T_{\rho}}{T}.
$$
Then using the approximation
\[
\frac{\tanh 
\sqrt{x^2+\rho_{ch}^2/4T^2}}{\sqrt{x^2+\rho_{ch}^2/4T^2}}\simeq 
\left\{ 
\begin{array}{ll} 
1-3^{-1}\left[x^2 +\rho_{ch}^2/4T^2\right], & x 
\leq 1; \\ x^{-1} -\rho_{ch}^2/8T^2x^3, & x>1, 
\end{array} \right.\] 
one directly comes to the expression needed:  
\begin{equation} 
\rho_{ch}(T)\simeq 
2.62T_{\rho}\sqrt{T_{\rho}/T-1}.\label{25} 
\end{equation} 
Recall that the generally accepted 3D result is 
$\Delta_{BCS}(T) =3.06T_{BCS}^{MF}\sqrt{T_{BCS}^{MF}/T-1}$ 
\cite{Abr1} and this small difference can be explained by the above 
approximation what, however, is suitable for the following below 
qualitative discussion (see next Section).  

The dependence (\ref{25}) has to be substituted in equation 
(\ref{19}).  And again because of $\mu 
/2T_{BKT}\simeq\epsilon_F/2T_{BKT}\gg 1$ and $\rho_{ch} 
(T_{BKT})/2T_{BKT}\ll 1$ when $T_{BKT}\rightarrow 
T_{\rho}$ this equation can be written as 
\begin{eqnarray} 
\frac{\epsilon_{F}}{4T_{BKT}}\left[1-\frac{\rho_{ch}^2}{4T_{BKT}^2}
\frac{\partial}{\partial (\rho_{ch}^2/4T_{BKT}^2)}\right]
\int_{0}^{\infty}dx\left(\frac{1}{\cosh^{2}x}-\right.
                                                   \nonumber \\
\left.\frac{1}{\cosh^{2}\sqrt{x^2+\rho_{ch}^2/4T_{BKT}^2}}\right)=1.
\label{26}
\end{eqnarray}
At last, using expansion in $\rho_{ch} /2T_{BKT}$ in integral 
(\ref{26}), the latter can be transformed to 
\begin{equation} 
\frac{a\epsilon_{F}}{8T_{BKT}}\left(\frac{\rho_{ch}}{2T_{BKT}}\right)^4=1,
\label{27} 
\end{equation} 
where the numerical constant
$$ a=\int_{0}^{\infty}dx\frac{\tanh^2x-x^{-1}\tanh x+1}
{2x^2\cosh x}\simeq 1.98.  
$$ 
Combining now (\ref{25}) and (\ref{27}) one comes to the final simple
relation between $T_{\rho_{ch}}$ and $T_{BKT}$ for the large carrier 
density:
$$ 
T_{BKT}\simeq 
T_{\rho}(1-1.17\sqrt{T_{\rho}/\epsilon_{F}}), 
$$ 
i.e. $T_{BKT}$ as a function of $n_f$ really approaches $T_{\rho}$
(or $T_{BCS}^{MF}$) (see fig.2).

As to crossover region defined by the equality $\mu\simeq 0$
it is easy to convince from the equations (\ref{18})-(\ref{21}) and
fig.2 that it corresponds to the densities when the temperatures $T_{\rho}$
are essentially different. It is important that because of relatively
low for the phonon case value of the bound pair-state energy
and so very small region of negative $n_{f}$ \cite{Lok2}, the 
behaviour $T_{BKT}(n_{f})\sim\epsilon_{F}$ hardly corresponds
to Bose-Einstein condensation and in fact takes place at $\mu >0$.

4) $g_{4F}<0, g_{ph}\not= 0$, but $g_{ph}^2>>|g_{4F}|$.

This condition provides the total fermion-fermion attraction channel 
only, so $\rho_{ch}\not=0$ (see below). As it was said above the 
local 4F-repulsion qualitatively can correspond to the screened 
Coulomb repulsion.  In this situation the cut parameter must be 
introduced to avoid the divergence in (\ref{20}). It is 
interesting to consider two situations:  i) the cut parameter (the 
boundary Coulomb frecuency $\omega_c$) goes to infinity (i.e.  
small concentration or local pairing) and ii) when $\omega_c$ is 
large but finite $\omega_c>>\omega_0$ but 
$g_{ph}>>|g_{4F}|$. Probably it is the 
situation what is intimately related to the real HTSCs 
(and superconductors at all).

i) The case 1) is restored in general but the
effective 4F-constant $g_{4F}^{**}<0$. It is important that $T_{BKT}$ 
preserves its linear assymptotics at small $n_{f}$.

ii) In this case $g_{4F}$ in (\ref{20}) can be substituted by 
$g_{4F}D(i\omega_n)$ with $\omega_0=\omega_c$ (the propagator $D$ is 
defined by (\ref{2})).  The situation is similar to the case 3) 
but the effective coupling constant is smaller. Such a decreasing 
leads to the narrowing of the abnormal normal phase region  
because of lowering of the temperature $T_{\rho}$. In particular,
it is not difficult to obtain the well known Tolmachev logarithmic 
correction to $T_{\rho}$ (see, for example, \cite{Von1}):  
$$ 
T_{\rho}=\frac{2\gamma\omega_0}{\pi}exp(-\frac{1}{g_{ph}^2m/2\pi 
-\mu^{*}}),
$$
where $\mu^{*}=g_{4F}N(0)/(1+g_{4F}N(0)\ln (\epsilon_F/\omega_0))$
($N(0)$ is the density of states at the Fermi surface).

\section{Conclusion}

The model proposed to describe the possible two-stage superconducting 
phase transition in 2D (and quasi-2D) metallic systems is in fact 
very simplified in order to investigate their most typical and 
general features. All the more surprisingly that it catches some 
essential details which are characteristic for underdoped HTSC copper 
oxides. In particular, the experimental data demonstrates 
\cite{Uem1,Uem2} that i) indeed for low $n_{f}$ the critical 
temperature $T_{c}$ is proportional to $n_{f}$ (what is simply
$\epsilon_{F}$), ii) $T_{c}$ shows saturation when $n_{f}$ approaches
so called "optimal doping" (i.e. carrier concentration when $T_{c}$
as function of $n_{f}$ reaches its highest possible in given compound 
value), iii) the ratio $T_c/\epsilon_F$ in these and other "exotic"
superconductors is as high as $10^{-2}-10^{-1}$ what independently 
points out on rather small Fermi energy, etc (for details see 
\cite{Uem2}).

One would think that the pecularities mentioned receive their natural 
explanation on the basis of the model of metal with indirect 
inter-fermion interaction if the temperature $T_{BKT}$ is implied
as critical one $T_{c}$ (this is justified for pure 2D systems 
\cite{McKen1}). In quasi-2D model because of the third spatial 
direction and the phase fluctuation stabilization the true 
temperature of ordinary homogeneous ordering arises 
\cite{Uem2,review} (see also \cite{Gor1}).

As regards the second temperature, $T_{\rho}$, it usually introduced 
by empirically as some temperature point $T^{*}$ where observable
spectral (or magnetic) properties of HTSC begin to deviate 
appreciably from their standard for normal metallic state behaviour
\cite{Lev1,Bat1,Ong1,Ber1,Pin1}. As a rule such a deviation is 
connected with appearance of fluctuating (short-living) pairs. We, 
however, showed that some finite number of these pairs does exist and 
is to formed at definite temperature due to phase transition between 
normal and pseudogap (also normal) phases. The only difference with 
supposed dependence $T^{*}$ on the density of doped holes consists of 
assymptotics at $n_{f}$ decreasing: we have obtained that this 
temperature is also reduced while usually (see, for example, 
\cite{Uem2}) $T^{*}$ is presumably depicted as such one that 
incriases with $n_{f}$ decreases. It seems that the latter has no 
satisfactory grounds.  Nevertheless it must be stressed that the 
above limit $T_{\rho}(n_{f}) \rightarrow 0$ when $n_{f}\rightarrow 0$ 
can not be also considered as sufficiently regular because of the 
growth here of the neutral OP fluctuations contributions of which was 
not taken into account and which become rather important at small
$n_{f}$.

At last the model under consideration qualitatively correctly 
describes the explicit narrowing of the pseudogap phase area
at carrier density increase (such a diminution results in rather
rapid reapproachment the temperature $T_{c} (T_{BKT})$ and $T_{\rho}$
and their experimental confluence (indistinguishableness) in BCS 
limit.

Some important problems remain open and are to be solved. Among them 
there are: more complete and deep development of the model which
has to consider different kinds of the dispersion laws for the 
intermediate bosons; more careful taking into account of the 
Coulombic repulsion; neutral OP fluctuations, especially at low 
$n_{f}$; generalization of the approach on the case of non-isotopic
pairing. On the other hand, high-$T_{c}$ compounds must be 
investigated in the frame of more realistic model that such their
pecularities as magnetism of cuprate layers, non-qudratic free 
carrier dispersion law with van Hove singularities in the hole 
density of states. One of the most interesting problem is to obtain
doping and temperature effective action which is equivalent to
Ginzburg-Landau potential because in many cases the phenomenology is 
more preferable.

We would like to thank Prof.~V.P.~Gusynin, Drs. S.G.~Sharapov and
I.A.~Shovkovy for extremely interesting and useful debates concerning 
the questions rised above. Expecially we are indebted I.A.~Shovkovy 
for reading the manuscript and comments.

\newpage
\begin{center}
Figures caption
\end{center}

Fig.1 \\ 
Phase $T$ -- $n_f$ diagram of 2D metal with 4F fermion 
attraction. The lines correspond to the functions 
$T_{\rho}(n_{f})$ (the upper line) and $T_{BKT(n_{f})}$ 
(the lower one) at $\lambda =0.5$.  The figures I, II and III show 
the regions of the normal, abnormal normal (pseudogap) and 
superconducting phases, respectively.

Fig.2 \\ 
Phase $T$ -- $n_f$ diagram of 2D metal with indirect inter-carrier
attraction. Similarly to the Fig.1, the lines correspond to the  
functions $T_{\rho}(n_{f})$ and $T_{BKT(n_{f})}$
and separate the same regions ($\lambda =0.5$). 

\end{document}